\documentclass[12pt]{article}
\usepackage{amsmath,amssymb,amsthm,amsfonts,color,colordvi,url}

\newcommand{\I}{\makebox[0pt][l]{1}\hspace{0.3ex}\mbox{1}}

\newcommand{\Tr}{\text{Tr}}

\newtheorem{theo}{Theorem}
\newtheorem{defi}{Definition}
\newtheorem*{rem}{Remark}
\theoremstyle{definition}

\newtheorem{propo}{Proposition}

\begin{document}

\title{Minimal Bell-Kochen-Specker proofs with POVMs on qubits}

\author{Andr\'e Allan M\'ethot\\[0.5cm]
\normalsize\sl D\'epartement d'informatique et de recherche op\'erationnelle\\[-0.1cm]
\normalsize\sl Universit\'e de Montr\'eal, C.P.~6128, Succ.\ Centre-Ville\\[-0.1cm]
\normalsize\sl Montr\'eal (QC), H3C 3J7~~\textsc{Canada}\\ 
\normalsize\url{methotan}\textsf{@}\url{iro.umontreal.ca} }

\maketitle
\begin{abstract}
There are many different definitions of what a Bell-Kochen-Specker
proof with POVMs might be. Here we present and discuss the minimal
proof on qubits for three of these definitions and show that they
are indeed minimal.
\end{abstract}

Einstein, Podolsky and Rosen argued in their 1935 paper that quantum
mechanics was not complete~\cite{epr35}. Their argument is based on
the fact that there seems to be some weird correlations on bipartite
measurements of a singlet state, which led Einstein to qualify this
phenomenon as ``spukhafte Vernwircklungen'' (spooky action at a
distance). In their opinion, this could only entail the existence of
another theory, a local realistic theory, that had variables which
where hidden to us. The ignorance of the hidden variables would lead
us to think that quantum mechanics is probabilistic and would ensure
the Heisenberg uncertainty principle. For almost 30 years, it was
debated as to whether such a theory existed and even as to whether
this question was a valid question for science---Pauli once compared
this question to how many angels can sit on a pin~\cite{bb71}. In
1964, Bell showed that, if the predictions of quantum mechanics are
correct and entanglement can resist space-like separation, no
realistic local hidden variable theory of quantum mechanics could
exist~\cite{bell64}. Independently, Bell~\cite{bell66} and Kochen
and Specker~\cite{ks67} showed that if such a theory existed, it had
to be contextual, independently of whether entanglement can resist
space-like separation. These proofs will be called BKS theorems
hereafter.

In Kochen-Specker paper, they proved that a three level state, or
qutrit, could not be described has a non-contextual realistic
theory. Such a non-con\-textual local-hidden-variable (LHV) theory
cannot assign a binary value (``yes''/``no'') to each element of a
measurement such that every von Neumann (projective) measurement on
a qutrit would have one and only one element with value ``yes'' and
that this value is independent of the measurement. In other words,
once the value ``yes''/``no'' is assigned, it stays the same for
this element in every measurement. Their proof is equivalent to a
non-two-coloring proof of a particular graph with 117 vertices where
every complete sub-graph of three vertices as one and only one
vertex of the first color. However, we know that their proof could
not be adapted to a two level state, or qubit, without some
modification~\cite{bell66}. In the original BKS proofs, Bell and
Kochen-Specker used von Neumann measurements. What if, to show the
non-contextuality of a qubit, one requires a more general form a
measurements? The most general form is detailed by the
positive-operator-valued measure formalism, henceforth called POVM.

We first give the definition of POVMs and contextuality. We then
give three definitions of BKS proofs with POVMs, each followed by
their minimal proof of the non-contextuality of the qubit and a
proof that it is in fact the minimal proof. It is to be noted that
there exist many more definitions of what a BKS proof with POVMs
might be, see for example~\cite{rob}, and that a consensus as to
which is rejects the class of LHV theories that was actually meant
by EPR as not been reached.

A POVM is a family of positive matrices $\{M_i\}$ such that $M_i=
A_i^{\dagger}A_i$ and $\sum M_i =\I$, where $M_i$ is called a POVM
element. On state $\rho$, a POVM will output $i$ with probability
$\Pr[i]=\text{Tr}(\rho M_i)$ with the state $\frac{1}{\text{Tr}(\rho
M_i)}A_i\rho A_i^{\dagger}$ as the quantum residue. When talking
about POVM elements, I will use the notation $\perp$ to denote the
inverse of the element: $M+M^{\perp}=\I$.

\begin{defi}
We say that a theory is contextual if the value of a physical
operator depends on the context in which it is measured.
\end{defi}
\begin{rem}
Quantum mechanics is \emph{not} a contextual theory.
\end{rem}
When we measure the state $\rho$ with a measurement $\{M_i\}$, we
get the result $i$ with probability $\Tr(\rho M_i)$, which does not
depend on the other elements of the measurement $M_j$ for $j\neq i$.

\begin{defi}
\label{simple} A BKS proof that quantum mechanics cannot be
described by a non-contextual realistic local variable theory can be
made if we consider that two measurement elements, which are
mathematically identical, are physically equivalent.
\end{defi}

\begin{propo}
\label{simplepropo} No local realistic non-contextual description
of a qubit exists according to Definition~\ref{simple}.
\end{propo}

\begin{proof}
In the POVM $\{\I/2,\I/2\}$, one cannot assign one and only one
``yes'' to an element per POVM since both elements are the
same~\cite{rob,toner}.
\end{proof}

\begin{theo}
The minimal proof of Proposition~\ref{simplepropo} requires one
POVM of two elements.
\end{theo}

\begin{proof}
To prove such a statement, we need to show that such a proof exists
and that a smaller one cannot be built. The proof was given above
and it is easy in this case to see that a smaller one could not
exist. A smaller proof would have one POVM of one element. For such
an ensemble, we can always assign one and only one ``yes'' per POVM
since the only POVM we have, $\{\I\}$, has only one element.
\end{proof}

\begin{defi}
\label{diff} A BKS proof that quantum mechanics cannot be described
by a non-contextual realistic local variable theory makes sense only
if each element in a measurement is considered distinct from the
other elements.
\end{defi}

\begin{propo}
\label{diffpropo} No local realistic non-contextual description of
a qubit exists according to Definition~\ref{diff}.
\end{propo}

\begin{proof}
The proof of this statement presented here can be found in Cabello's
paper \cite{cabello02} and is due to Nakamura. If we consider the
following POVMs:
\begin{equation}\label{eq:cabello}
\begin{aligned}
& \{A/2,A^{\perp}/2,B/2, B^{\perp}/2\},\\
& \{A/2,A^{\perp}/2,C/2,C^{\perp}/2\}\ \text{and}\\
& \{B/2,B^{\perp}/2,C/2,C^{\perp}/2\},
\end{aligned}
\end{equation}
since every element appears twice and the number of ``yes'' needed
is odd, 3, one cannot assign non-contextually one and only one
``yes'' per POVM.
\end{proof}

\begin{theo}
\label{difftheo} The minimal proof of Proposition~\ref{diffpropo}
requires three POVMs of four elements each.
\end{theo}

\begin{proof}
First let us consider reducing the number of POVMs. If we only have
one POVM with distinct elements, it is easy to assign non-contextual
``yes''/``no'' values to the elements, we choose any element at
random to have the ``es'' value and give the ``no'' value to all the
other elements. For two POVMs, we can select any POVM element that
appears twice, or any two POVM elements that appear once, one in the
first POVM and the other in the second, to give the ``yes'' value.
It is therefore clear that we need at least three POVMs.

So far we have established that we need three POVMs and that three
POVMs with four elements are sufficient. Now let us describe what
happens if we reduce the number of elements. If we consider POVMs,
with only one element, then it is always the same POVM with the same
element, hence it is easy to have a non-contextual description of
what happens. POVMs composed of two elements are always composed of
an element $M$ and its \emph{unique} inverse $M^{\perp}$. It is
therefore impossible to construct a proof of contextuality.

Let us now examine the case of three POVMs of three elements each.
We can express them, without loss of generality, in the form of
$\{A_1,B_1,C_1\}$, $\{A_2,B_2,C_2\}$ and $\{A_3,B_3,C_3\}$, such
that $A_i\neq B_j$ and $A_i\neq C_j$ for all $(i,j)$. Let us now
turn our attention on $A_1$, $A_2$ and $A_3$. Either (1) all these
elements are all the same, either (2) two elements of the three
elements are the same or either (3) they are all different. In all
the three cases we can simply assigne the ``yes'' value to the $A_i$
elements and the ``no'' value to the other elements without running
into a contradiction.

This as established the we need at least three POVMs, three elements
in each POVM and in the special case where we have exactly three
POVMs, we require four elements in each POVM. A possibility yet
explored would be to construct a proof using four POVMS of three
elements each, yielding the same total number of elements. However,
such a construction is impossible. Let us write the ensemble in the
form $\{A_1,B_1,C_1\}$, $\{A_2,B_2,C_2\}$, $\{A_3,B_3,C_3\}$ and
$\{A_4,B_4,C_4\}$. We can always relabel the elements such that
either $A_i\neq B_j$ and $A_i\neq C_j$ for all $(i,j)$, or in a way
that is isomorphic to the ensemble $\{A_1,B_1,C_1\}$,
$\{A_1,B_2,C_2\}$, $\{A_1,B_3,C_3\}$ and $\{B_1,B_2,B_3\}$. In both
cases, it is easy to non-contextually assign ``yes'' values without
creating a contradiction.
\end{proof}

\begin{defi}
\label{nopropor} A BKS proof that quantum mechanics cannot be
described by a non-contextual realistic local variable theory must
be formulates such that measurements elements $M_i$ of the same POVM
are not proportional to one another, $M_j \neq \gamma M_i$ for $j
\neq i$ and $\gamma > 0$.
\end{defi}

\begin{propo}
\label{noporporpropo} No local realistic non-contextual
description of a qubit exists according to
Definition~\ref{nopropor}.
\end{propo}

\begin{proof}
We can use the Cabello's ensemble given in the
Equation~\eqref{eq:cabello} as we did for
Proposition~\ref{diffpropo}.
\end{proof}

\begin{theo}
The minimal proof of Proposition~\ref{noporporpropo} requires
three POVMs of four elements each.
\end{theo}

\begin{proof}
The proof is exactly the same as the proof of
Theorem~\ref{difftheo}.
\end{proof}

\begin{defi}
\label{ben} A BKS proof that quantum mechanics cannot be described
by a non-contextual realistic local variable theory has to assign
``yes''/``no'' values only to the elements of the POVM that could
not appear twice in any POVM, $\vert M_i \vert > \frac{1}{2}$.
\end{defi}

Toner, Bacon and Ben-Or~\cite{toner} have a proof that one cannot
have a non-contextual description of a qubit according to
Definition~\ref{ben}. Their proof has $9$ POVMs with $3$ to $4$
elements each for total of $31$ elements.

Although the minimal proofs of Proposition~\ref{diffpropo} and
Proposition~\ref{noporporpropo} are the same, it must be noted that
the definitions that led to these proof are fundamentally different.
Where Definition~\ref{diff} only requires the elements to be
mathematically distinct for them to be physically different,
Definition~\ref{nopropor} states that output direction distinguishes
physically different measurement elements.

Further work needs to be done to establish what are the minimal
proofs for the definition given in~\cite{toner}, but there is a far
more fundamental open question. It is not clear which definition of
a BKS proof with POVMs is the most natural one. We could argue that
Definition~\ref{simple} is the correct one, that it is simple to
show the that quantum mechanics cannot be described by a realistic
non-contextual theory and that God indeed plays dice, i.e. Nature is
not deterministic. Some might argue that it is evident that using
the same elements in a POVM many times will lead to a contradiction
and thus Nature assigns indices to elements that appear twice or
more in order to make them distinct, in which case something more
refined must be used. One can present such arguments for every
definition presented here. So the question remains: What is the
definition of a BKS proof with POVMs that is more natural? Maybe
such a question is to be left to philosophers, but as Bell, Kochen
and Specker showed, one should not make such a claim hastily: there
could be some yet unforseen consequence to adopt the \emph{point de
vue} of one of the definitions given here.

Since we know that there is a strong link between the conventional
BKS proofs and pseudo-telepathy~\cite{aravind99,hr83}, often wrongly
called a Bell theorem without inequalities~\cite{methot05}, one
might wonder why we cannot transform a BKS proof with POVMs on a
qubit into a pseudo-telepathy game on an entangled qubit pair
\cite{bmt04}.

The author is particularly in dept to Ben Toner for sharing is
thoughts and knowledge on the subject and would like to thank
Gilles Brassard for helpful discussions.

\end{document}